\documentclass[11pt]{article}
\pdfoutput=1 % if your are submitting a pdflatex (i.e. if you have images in pdf, png or jpg format)
\usepackage{./aug/jheppub}

\usepackage{graphicx}
\usepackage{amsfonts,amsmath,amssymb}
\usepackage{tikz}
\usepackage{longtable}
\usepackage{verbatim}
\usepackage{array,setspace,mathrsfs,yfonts,dsfont,bbm,colonequals,amscd}
\usepackage{relsize,suffix,mathtools,cancel,bbm}

\usepackage{multirow}
\usepackage{longtable}

\newcommand{\be}{\begin{eqnarray}}
\newcommand{\ee}{\end{eqnarray}}
\newcommand{\bea}{\begin{eqnarray}}
\newcommand{\eea}{\end{eqnarray}}

\newcommand{\bn}{\begin{enumerate}}
\newcommand{\en}{\end{enumerate}}

\def\dim{{\rm dim}}

\title{A Nilpotency Index of Conformal Manifolds}

\preprint{}

\author[a]{Zohar Komargodski,\!}
\author[b]{Shlomo S. Razamat,\!}
\author[b]{Orr Sela,\!}
\author[c]{and Adar Sharon}

\affiliation[a]{Simons Center for Geometry and Physics, Stony Brook, New York, USA}
\affiliation[b]{Department of Physics, Technion, Haifa, 32000, Israel}
\affiliation[c]{Department of Particle Physics and Astrophysics, Weizmann Institute of Science,\\ Rehovot 7610001, Israel}

\emailAdd{zkomargodski@scgp.stonybrook.edu}
\emailAdd{razamat@physics.technion.ac.il}
\emailAdd{sorrsela@campus.technion.ac.il}
\emailAdd{adar.sharon@weizmann.ac.il}

\abstract
{
We show that  exactly marginal operators of Supersymmetric Conformal Field Theories (SCFTs) with four supercharges cannot obtain a vacuum expectation value at a generic point on the conformal manifold. 
Exactly marginal operators are therefore nilpotent in the chiral ring. This allows us to associate an integer to the conformal manifold, which we call the nilpotency index of the conformal manifold.  We discuss several examples in diverse dimensions where we demonstrate these facts and compute the nilpotency index. 
}

\newcommand\mc[1]{\mathcal{#1}}

\begin{document} 
	
\maketitle
\flushbottom

%%%%%%%%%%%%%%%%%%%%%%%%%%%%%%%%

\section{Introduction}
\label{int}

The existence of continuous families of Conformal Field Theories (CFTs) is a surprising  phenomenon that occurs mostly in supersymmetric theories. Indeed, without supersymmetry, in two dimensions it is not expected that there are continuous families of conformal field theories except in the case of $c=1$~\cite{Cardy:1987vr}. In higher dimensions, no examples are known (with any finite central charge) of non-trivial conformal manifolds. Given a marginal primary operator, for it to be exactly marginal infinitely many intricate constraints need to be satisfied (see \cite{Gaberdiel:2008fn,Komargodski:2016auf,Bashmakov:2017rko,Behan:2017mwi} for the first couple of constraints) and it is quite hard to believe that such examples exist without supersymmetry. 

In this note we study some canonical questions about conformal manifolds that occur in supersymmetric theories with four supercharges (or more). The existence of such conformal manifolds has a long history, see {\it e.g}  \cite{Sohnius:1981sn,Howe:1983wj,Parkes:1984dh}, and a plethora of examples was first systematically constructed in \cite{Leigh:1995ep} and later in \cite{Green:2010da} (see also~\cite{Kol:2010ub}). Let $P$ be such an SCFT. For convenience, the reader may keep in mind either $\mathcal{N}=2$ theories in 3d or $\mathcal{N}=1$ theories in 4d. An exactly marginal operator resides in a multiplet of a chiral primary operator $\mathcal{O}_I$ of dimension $d-1$. The subscript $I$ here ranges over a basis of the space of exactly marginal operators. If we deform the action by $$\int d^dx h^I Q^2\mathcal{O}_I+c.c.\,,$$ (with $Q$ standing for the supercharge) then the theory remains conformal. We thus get a space of SCFTs with local coordinates $h^I$. This space is called the conformal manifold $\mathcal{M}_c$, and an exactly marginal operator $\mathcal{O}$ in a generic direction can be expressed as
\begin{equation}
\label{genO}
\mathcal{O}=\sum_{I=1}^{\dim\mathcal{M}_{c}}h^{I}\,\mathcal{O}_{I}\,.
\end{equation}

General considerations show that the conformal manifold is endowed with the structure of Riemannian geometry~\cite{Kutasov:1988xb}. In supersymmetric theories this space is furthermore endowed with a complex structure and a K\"ahler metric \cite{Seiberg:1988pf,Asnin:2009xx}. We will often refer to $\mathcal{O}$ as the exactly marginal operator even though the actual exactly marginal operator is $Q^2\mathcal{O}$. 

Supersymmetric theories with four (or more) supercharges, whether they are conformal or not, admit a chiral ring. The chiral ring is defined as the space of all local operators satisfying $\overline Q \mathcal{O}=0$ with the identification that if $\mathcal{\widetilde{O}}-\mathcal{O}=\overline{Q}V$, with $V$ any local operator, then $\mathcal{\widetilde{O}}$, $\mathcal{O}$ are deemed equivalent in the chiral ring, $\mathcal{\widetilde{O}}\sim {\cal O}$. The ring structure is obtained by multiplying chiral ring operators with each other. Since there cannot be a singularity in the product of two chiral ring operators, the product is well defined (it may or may not be zero).
The chiral primary operators $\mathcal{O}_I$ corresponding to exactly marginal deformations are of course members of this chiral ring. 

For every $P\in \mathcal{M}_c$ there is a chiral ring and the chiral ring depends on $P$ in a potentially nontrivial way. For instance, the multiplication rules in the ring may change as a function of $P$ and also chiral ring elements may appear and disappear as we change $P$. Since we need a connection to compare the chiral rings of two nearby fixed points, the chiral ring is a bundle over $\mathcal{M}_c$.

Therefore, the exactly marginal operators  $\mathcal{O}_I$ play two roles -- they are the operators that are used to move along $\mathcal{M}_c$ but at the same time at every given point on $\mathcal{M}_c$ they are elements of the chiral ring. Loosely speaking, one can say that the chiral ring bundle contains the tangent bundle. Therefore a natural question is what are the chiral ring relations obeyed by products of the $\mathcal{O}_I$. 

Here we will argue that  apart from complex co-dimension 1 (or higher) subspaces, the $\mathcal{O}_I$ are everywhere nilpotent. Namely, there is some integer for which 
\begin{equation}
\label{Nil}
\mathcal{O}_I^{k} \sim 0\,.
\end{equation}
A priori, $k$ may depend on our position in the conformal manifold $\mathcal{M}_c$ and on $I$. However, in theories with at least four supercharges, due to  holomorphicity properties, the index $k$ can only ``jump" along complex co-dimension 1 (or higher) subspaces of the conformal manifold, and so we should be able to bypass these singularities and define an index which is constant away from these subspaces. Similarly, $k$ can only jump along complex co-dimension 1 (or higher) directions of $\mathcal{M}_c$, and so it should be identical for generic operators of the form \eqref{genO} ({\it i.e.} for generic $h^I$). 
This endows the conformal manifold with a global integer invariant $k(\mathcal{M}_c)$ which is the nilpotency index of an exactly marginal operator corresponding to a generic direction at a generic point of the conformal manifold (away from co-dimension 1 (or higher) singularities).

Let us now list some comments and applications:

\begin{itemize}
	\item Sometimes $\mathcal{M}_c$ is conjectured to have several weak coupling corners.  (For a surprising recent class of examples see~\cite{Razamat:2019vfd,Razamat:2020gcc,Kim:2018bpg}.) Then we can compute $k(\mathcal{M}_c)$ in several different ways and perform a nontrivial consistency check of such dualities. We hope to return to this in the future.
	
	\item The chiral ring at $P\in \mathcal{M}_c$ is closely related to the space of supersymmetric vacua of the theory $P$. More precisely, we should discard the nilpotent operators from the chiral ring and what remains is isomorphic to the space of holomorphic functions over the supersymmetric vacua. 
	In one direction this correspondence is clear: the chiral ring relations must be obeyed by the expectation values of chiral operators in supersymmetric vacua. The other direction is that supersymmetric vacua must always yield corresponding local chiral operators which can attain expectation values at the vacua. This is not at all obvious. It was proven in Lagrangian theories~\cite{Luty:1995sd} and remains conjectural in general. We will assume this correspondence between the chiral ring and the space of vacua to be true. Then to prove~\eqref{Nil} it is sufficient to prove that there are no supersymmetric vacua where $\langle\mathcal{O}_I\rangle\neq 0$.
	
	\item We can consider, for instance, $\mathcal{N}=2$ theories in four dimensions viewed as $\mathcal{N}=1$ theories. If we consider the conformal sub-manifold of $\mathcal{N}=2$ preserving exactly marginal deformations then we can show that within this sub-manifold there are no complex co-dimension 1 loci where the nilpotency index diverges. We may define an integer index $k(\mathcal{M}_c)$ for the $\mathcal{N}=2$ sub-manifold. This integer may or may not agree with the integer of the whole conformal manifold since the sub-manifold of $\mathcal{N}=2$ preserving deformations may be a singular locus inside $\mathcal{M}_c$. 
	
	\item In 2d theories all the operators in the chiral ring are nilpotent and in particular our result~\eqref{Nil} is trivially satisfied. Yet it is still interesting to find the exponent $k(\mathcal{M}_c)$ in 2d models. We discuss some examples in the main text. Another peculiarity of 2d is that the conformal manifold is in fact a direct product of the chiral and twisted chiral deformations. Therefore there are in principle two integer indices to discuss.
	
	\item The superconformal index \cite{Romelsberger:2005eg,Kinney:2005ej,Dolan:2008qi} (for a review see~\cite{Rastelli:2016tbz}) provides a wealth of information  about BPS operators in SCFTs. The extraction of the integer $k(\mathcal{M}_c)$ however seems nontrivial, if possible at all. Instead, we rely on some weak coupling techniques in order to extract the integer $k(\mathcal{M}_c)$. 
	
\end{itemize}

The structure of the paper is as follows.
In section 2 we present an argument that the exactly marginal operators $\mathcal{O}_I$ are nilpotent aside from complex co-dimension 1 (or higher) loci. In section 3 we discuss some simple examples where we either illustrate our theorem or compute the nilpotency index.

\

\section{Vanishing Expectation Values}
\label{thrm}

\subsection{Elementary Results}\label{elementary}

In this subsection we would like to review a few important facts about supersymmetric theories with four supercharges and an $R$-symmetry. The discussion below will be entirely in the realm of classical physics of supersymmetric theories with $n$ chiral superfields $\Phi_i$, each carrying $R$-charge $q_i$. Conformal invariance will not be used in this subsection.

Suppose the $R$-symmetry is spontaneously broken. Then we can isolate an $R$-axion superfield $\Psi$ such that $e^{i\Psi}$ has $R$-charge 2 and the superpotential can be written as 
$$W=e^{i\Psi} F(X_i)~,$$
where the $X_i$ are R-neutral superfields and $F$ is an arbitrary function. So far we have only assumed that the $R$-symmetry is spontaneously broken but we did not assume that the vacuum is supersymmetric. 

Nelson-Seiberg \cite{Nelson:1993nf} argued that if $F$ is a ``generic" function then no SUSY vacua will exist since the critical points of $W$ require $F=\partial_iF=0$ which gives one more equations than variables in $F$ and hence in general no solution is expected to exist.

Of course, since in supersymmetric theories the superpotential is not renormalized, $F$ does not have to be generic~\cite{Grisaru:1979wc,Seiberg:1993vc} and there are many examples with SUSY vacua which spontaneously break the $R$-symmetry (the maximally supersymmetric theory being one such example).

Here we will be interested in a variant of the statement which applies rigorously without having to require genericity of $F$. Let $F$ be an arbitrary function which depends analytically also on some collection of parameters $\epsilon^A$, $$F=F(X_i; \epsilon^A)~.$$
Let us assume that the equations $F=\partial_iF=0$ admit solutions ({\it i.e.} SUSY vacua) which depend analytically on $\epsilon^A$. Then we would like to prove that \begin{equation}\label{Lemma}\frac{\partial}{\partial \epsilon^A}W\biggr|_{SUSY}=0~\,,\end{equation}
where $\biggr|_{SUSY}$ means that we are evaluating the partial derivatives on the SUSY vacua.

The proof proceeds by parameterizing the SUSY solutions by $X_i(\sigma;\epsilon^A)$ where $\sigma$ is some collection of parameters. It must be true that $F(X_i(\sigma;\epsilon^A);\epsilon^A)=0$ on the SUSY vacua (there is no summation over $A$). Now taking a derivative with respect to $\epsilon^A$ we find 
$$0={d\over d\epsilon^A} F = \partial_i F{\partial X_i \over \partial \epsilon^A}+{\partial F\over \partial \epsilon^A}~. $$
Since at SUSY vacua $\partial_i F=0$, we remain with the desired conclusion ${\partial F\over \partial \epsilon^A} \biggr|_{SUSY}=0$ (this is equivalent to~\eqref{Lemma}). Soon it will be clear why this conclusion is important. 

While our arguments here were in the realm of classical field theory, the conclusion holds more generally. In particular, the condition that SUSY vacua with broken $R$-symmetry require $F=0$ can be viewed as a special case of the inequality~\cite{Dine:2009sw}, which holds non-perturbatively. 

A key assumption above was that the space of solutions is an analytic function of the $\epsilon^A$ so that ${\partial X_i(\sigma;\epsilon^A) \over \partial \epsilon^A}$ makes sense. It is worth looking more closely into this innocuous-looking assumption. Consider for example the following classical superpotential:\footnote{We remind that in this subsection no reference to conformality is made.}
$$W=-XYZ+{1\over 3}a X^3+{1\over 3}b Y^3+{1\over 3}cZ^3~.$$
The critical point equations are 
$$XY=c Z^2~,\quad XZ=b Y^2~,\quad YZ=a X^2~.$$
We will be interested in analyzing the space of solutions in a neighborhood of $a=b=c=0$. Let us list all the cases (up to permutations): 
\begin{itemize}
\item $a=b=c=0$: Here we have three lines of SUSY vacua, parameterized by one of $X,Y,Z$ with the other two vanishing. So we have three complex lines of vacua meeting at one point. 
\item $a=b=0, c\neq0$: Here we must have $Z=0$ while exactly one of $X,Y$ may be nonzero. Thus we have two complex lines of vacua meeting at one point. 
\item $a=0, b,c\neq0$: Here we must have $Y=Z=0$ while $X$ may be nonzero. Thus we have a complex line of vacua. 
\item $a, b,c\neq0$: There is a complex line of vacua only if $abc=1$. Otherwise the only solution is $X=Y=Z=0$. However $abc=1$ does not pass near a small neighborhood of the origin in parameter space and hence it can be ignored. 
\end{itemize}

This example is quite educational. First, we see that it is not true that one can parameterize the SUSY vacua analytically in a neighborhood of $a=b=c=0$ as there are co-dimension 1, 2, and 3 loci with SUSY vacua that do not appear elsewhere. 
On the other hand, away from these loci, since the only solution is $X=Y=Z=0$, we find that as predicted by~\eqref{Lemma} it is indeed true that $\langle X^3\rangle =0 $, $\langle Y^3\rangle =0 $, $\langle Z^3\rangle =0 $.  
Furthermore, if one sits on the sub-manifold $a=0$ with generic small $b,c$ the theorem still holds for the respective operators which are activated in the action $\langle Y^3\rangle =0 $, $\langle Z^3\rangle =0 $.

In summary, the general result~\eqref{Lemma} applies away from potential complex co-dimension 1 (or higher) sub-manifolds, where the space of solutions may jump. But one can use~\eqref{Lemma} also inside those loci as the space of solutions within the loci itself obeys the same rules.
We will see applications of these results later.

\subsection{The Main Argument}

Let us carefully define the setup of our result. 
We consider a superconformal field theory in $d=3$ or $d=4$ dimensions with at least four supercharges (we will be using only four supercharges for the statement of the result). We assume that the theory  has a  conformal manifold ${\cal M}_c$ parametrized by  exactly marginal deformations $\left\{ \mathcal{O}_{I}\right\}_{I\in\{1\dots \text{dim}{\cal M}_c\}}$.
These operators carry $R$-charge 2 and dimension $\Delta=d-1$.

Let $P\in{\cal M}_c$ be some SCFT. Our primary interest is in the moduli space of supersymmetric vacua at $P$. We want to know something about the expectation values of the exactly marginal operators, $\left\langle \mathcal{O}_{I}\right\rangle$. The theory at $P$ may have a non-trivial moduli space of vacua with a particular vacuum which is distinguished, at the ``origin.'' This vacuum at the origin is distinguished by it not breaking spontaneously the conformal symmetry nor the $R$-symmetry.

Let us suppose that on the moduli space of vacua at $P$ there are indeed some vacua where $\left\langle \mathcal{O}_{I}\right\rangle$ is non-vanishing. These vacua therefore necessarily break the conformal symmetry and the $R$-symmetry, and have a massless $R$-axion and dilaton particles (along with their superpartners and, potentially, other matter). These vacua are clearly away from the origin.
According to this assumption, the $\mathcal{O}_I$ cannot be nilpotent chiral ring operators at $P$, for if they were nilpotent, they could not obtain a nonzero vacuum expectation value in any supersymmetric vacuum.

Let us now explore what happens as we study the theories around $P$, say the theory $P+\delta P$ which is given by deforming the superpotential by $W=W_P+\sum h^I\mathcal{O}_I$ with small $h^I$. As shown in the previous subsection, away from possible complex co-dimension 1 (or higher) subspaces, from~\eqref{Lemma} it follows that $\langle\mathcal{O}_I\rangle=0$ for all the supersymmetric vacua in a neighborhood of $h^I=0$.

As a result, at such generic points in a neighborhood of $P$ the operators $\mathcal{O}_I$ are nilpotent. (The operators $\mathcal{O}_I$ are chiral for all $h^I$ since they are exactly marginal. They are furthermore chiral primaries.) Therefore, what we have argued is that even if there is a point where the exactly marginal operators are not nilpotent, they must be nilpotent in a neighborhood of that point except possibly on some co-dimension 1 (or higher) loci. Loosely speaking, we see that the tangent space to the conformal manifold is a nilpotent element of the chiral ring bundle.

An operator $\mathcal{O}$ is said to be nilpotent if there is some power $k$ for which $\mathcal{O}^k \sim 0$. The smallest such integer $k$ is what we would call the ``nilpotency index.'' We have argued that all the exactly marginal operators $\mathcal{O}_I$ are nilpotent at generic points on the conformal manifold.
Since we argued that the nilpotency index may jump only on complex co-dimension 1 (or higher) submanifolds, such singularities may be avoided and we are therefore led to assign a nilpotency index for the whole conformal manifold $k({\cal M}_c)$.

Furthermore, the singular loci could be separately interesting as explained in the previous subsection. We can analyze them ignoring the other exactly marginal deformations and assign nilpotency indices to these loci. 
We will see examples of these ideas below.

\begin{figure}
	%[htbp ]
	\center\includegraphics[scale=0.8]{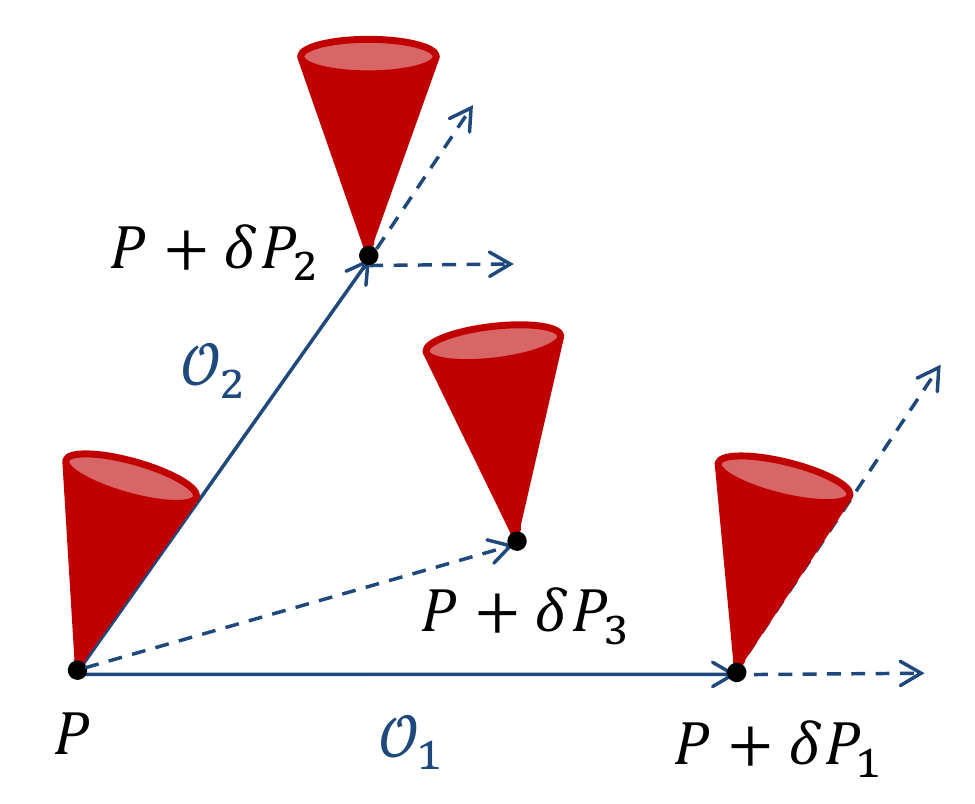}
	\caption{A depiction of the theorem. We start at a point $P$ of ${\cal M}_c$ with a set of exactly marginal deformations, ${\cal O}_1$ and ${\cal O}_2$.  The moduli space, depicted as a red cone, at $P$ might have a common direction with ${\cal M}_c$, say ${\cal O}_2$. However, once we move along ${\cal O}_2$ to $P+\delta P_2$ the moduli space cannot have a component along the ${\cal O}_2$ direction. Moreover, on a generic locus $P+\delta P_3$ of ${\cal M}_c$ the moduli space and ${\cal M}_c$ have no common directions. On complex co-dimension one (or higher) loci $P+\delta P_1$ some marginal operators, say ${\cal O}_1$ in the figure,  might have common directions.	}
	\label{theo}
\end{figure}

\

\section{Simple Examples}
\label{triv}

Let us now discuss several concrete examples in which our theorem can be easily seen to hold.

\

\subsection{$2d$ $\mathcal{N}=(2,2)$ Theories}
\label{2d}

The proof of the theorem that exactly marginal deformations are generically nilpotent relied on the analysis of moduli spaces of vacua. The theorem is trivially valid also in two spacetime dimensions since there are no moduli spaces~\cite{Coleman:1973ci} and hence all the operators, not just the exactly marginal ones, are nilpotent. 
It is still interesting to compute the nilpotency index of the operators which are exactly marginal, which is the aim of what follows.

Recall that for SCFTs in two dimensions with $\mathcal{N}=(2,2)$ supersymmetry, the $R$-charge of chiral primary operators\footnote{Similar comments apply for the axial $R$-symmetry and twisted chiral operators.} is bounded by the central charge $c$ as follows (see {\it e.g.} \cite{Lerche:1989uy} and following their notation): 
\begin{equation}
\label{eq:bound_on_R_charge}
q_R\leq \frac{c}{3}\,.
\end{equation}
This, in turn, results in a finite chiral ring, with all its elements being nilpotent. In particular, every exactly marginal operators is nilpotent with an index $k$ bounded by \begin{equation}
\label{eq:bound_on_index}
k\leq\frac{c}{3}+1\,.
\end{equation}

\subsubsection*{Landau-Ginzburg models with degree $n$ superpotential}

Let us consider $N\geq3$ copies of the $2d$ Landau-Ginzburg model of a single chiral superfield $X$ with a degree $N$ superpotential. That is, we examine the theory of $N\geq3$ fields $X_k$ ($k=1,\dots,N$) and the superpotential  
\begin{equation}
W=\frac{1}{N}\sum_{k=1}^{N}X_{k}^{N}\,.
\end{equation}
(The cases $N=1,2$ are excluded by virtue of them being empty SCFTs.)
In the infrared, this theory flows to $N$ copies of the $\mc{N}=(2,2)$ $A_{N-1}$ minimal model (see {\it e.g.} \cite{Witten:1993jg} and references therein). For $N>2$, there are generally many marginal operators with which we can deform the theory. In $\mc{N}=(2,2)$ SCFTs every marginal operator is exactly marginal, see~\cite{Bertolini:2014ela} and references therein for more details. For simplicity, we start by concentrating on a single direction in the conformal manifold corresponding to the exactly marginal operator  
\begin{equation}
X_1\cdots X_{N}\,,
\end{equation}
which exists for all $N\geq 3$. As a result, we have the superpotential 
\begin{equation}
\label{W2dn}
W=h\,X_1\cdots X_{N}+\frac{1}{N}\sum_{k=1}^{N}X_{k}^{N}\,,
\end{equation}
and the corresponding $F$-term relations are 
\begin{equation}
\label{F2dn}
h\,X_2\cdots X_{N}+X_1^{N-1}\sim0~,
\end{equation}
and analogous relations arise from the $F$-term equations for $X_2,\dots,X_N$ which are obtained via a cyclic permutation of the $X_i$'s. 
We are working in a scheme in which \eqref{W2dn} is not renormalized. 

In order to identify the various operators in the chiral ring, we will keep track of their charges under the discrete symmetries. There are $N-1$ independent $\mathbb{Z}_{N}$ symmetries which can be chosen as follows: 
\begin{equation*}
\left(X_1,\ldots,X_{N}\right)\rightarrow\left(e^{\frac{2\pi i}{N}}X_1,X_2,\cdots,X_{N-1},e^{\frac{2\pi i\left(N-1\right)}{N}}X_{N}\right)\,,
\end{equation*}
\begin{equation*}
\vdots
\end{equation*}
\begin{equation}
\left(X_1,\ldots,X_{N}\right)\rightarrow\left(X_1,X_2,\cdots,e^{\frac{2\pi i}{N}}X_{N-1},e^{\frac{2\pi i\left(N-1\right)}{N}}X_{N}\right)~.
\end{equation}

Let us denote by $\alpha_{j}$ ($j=1,\ldots,N-1$) the corresponding fugacities (which are some $N$-th roots of unity).
%\begin{equation}
%\alpha_{j}=e^{\frac{2\pi i}{n}}\,,\,\,\,\left(j=1,\ldots,n-1\right).
%\end{equation}
Then, the generating function of chiral operators, {\it a.k.a} Hilbert series (see {\it e.g.} \cite{Benvenuti:2006qr}), refined under these $\mathbb{Z}_{N}$ symmetries, is given by 
\begin{equation}
H=\textrm{Tr}\,t^{R}\prod_{j=1}^{N-1}\alpha_{j}^{Q_j}~.
\end{equation}
$Q_j$ stand for the charge under the $j$th $\mathbb{Z}_{N}$ symmetry, $R$ are the R-charges, and we trace over the chiral ring. This charge is defined mod $N$ but since the $\alpha_j$ are $N$th roots of unity, the Hilbert series is well defined. 
It evaluates to 
\begin{equation*}
H=\frac{\left(1-t^{\frac{2\left(N-1\right)}{N}}\alpha_1^{N-1}\right)\cdots\left(1-t^{\frac{2(N-1)}{N}}\alpha_{N-1}^{N-1}\right)\left(1-t^{\frac{2\left(N-1\right)}{N}}\alpha_1\cdots\alpha_{N-1}\right)}{\left(1-t^{\frac{2}{N}}\alpha_1\right)\cdots\left(1-t^{\frac{2}{N}}\alpha_{N-1}\right)\left(1-t^{\frac{2}{N}}\alpha_1^{N-1}\cdots\alpha_{N-1}^{N-1}\right)}
\end{equation*}
\begin{equation}
\label{Hn}
=1+t^{\frac{2}{N}}\left(\sum_{j=1}^{N-1}\alpha_{j}+\alpha_1^{N-1}\cdots\alpha_{N-1}^{N-1}\right)+\ldots+t^{2\left(N-2\right)}\,.
\end{equation}
The denominator corresponds to the generators of the ring and the numerators comes from the $N$ $F$-term relations.
At the lowest nontrivial order in this expansion, $2/N$, we can identify the  operators $X_k$ according to their charges: they all have an $R$-charge $2/N$ (following from the superpotential \eqref{W2dn}) and the $j$th $\mathbb{Z}_{N}$ charge $\delta_{jk}$ for $k=1,\ldots,N-1$ and $N-1$ for $k=N$. At order $t^2$ in the expansion, there are generally many (exactly) marginal operators, where only one is a singlet under all of the symmetries. The series truncates after $R$-charge $2(N-2)$, and so the maximal nilpotency index for a marginal operator is $N-1$, which agrees with the bound \eqref{eq:bound_on_index} when plugging in the central charge for these theories, $c=3N(1-\frac{2}{N})$ \cite{Lerche:1989uy}.

Importantly, at the maximal $R$-charge $2\left(N-2\right)$ we find a single operator
\begin{equation}
\label{maxordn}
\left(X_1\cdots X_{N}\right)^{N-2}\,,
\end{equation}
which is neutral under the discrete symmetries and can also be written in various different ways using the relations \eqref{F2dn}. Indeed, as shown in \cite{Lerche:1989uy}, there always exists a unique chiral operator which saturates the bound  \eqref{eq:bound_on_R_charge}. Note that for $N>3$, \eqref{maxordn} can be expressed as the $(N-2)$-th power of more than one marginal operator (taking into account the relations \eqref{F2dn}), as will be demonstrated below. Therefore, we see that on the line on the conformal manifold parameterized by $h$ there are exactly marginal operators such as $X_1\cdots X_{N}$ that have a nilpotency index $N-1$, and other exactly marginal operators that have a lower nilpotency index. In the Hilbert series we do not observe their $N-2$th power as it is zero in the chiral ring due to the $F$-term relations. 

We now consider the cases $N=3$ and $N=4$ in more detail, followed by a discussion for general $N$. For $N=3$, the Hilbert series $H$ is given by 
\begin{equation}
H^{(N=3)}=1+t^{\frac{2}{3}}\left(\alpha_1+\alpha_2+\alpha_1^2\alpha_2^2\right)+t^{\frac{4}{3}}\left(\alpha_1^2+\alpha_2^2+\alpha_1\alpha_2\right)+t^2\,,
\end{equation}
and the different contributions are as follows. At $R$-charge zero we have the identity operator (corresponding to the vacuum state) and at order $2/3$ the fundamental fields $X_1$, $X_2$ and $X_3$. At order $4/3$ we have $X_1^2$, $X_2^2$ and $X_3^2$, or alternatively (using the relations) $X_2X_3$, $X_1X_3$ and $X_1X_2$. Finally, at the maximal order 2 we have the single marginal operator 
\begin{equation}
X_1X_2X_3\sim-h^{-1}X_1^{3}\sim-h^{-1}X_2^{3}\sim-h^{-1}X_3^{3}\,,
\end{equation}
while the other inequivalent operators (of the form $X_{k}X_{m}^2$) vanish in the chiral ring. We therefore see that the nilpotency index of the (single) exactly marginal operator $X_1X_2X_3$ is 2. 

Moving to $N=4$, the series $H$ is given by 
\begin{equation}
H^{(N=4)}=1+t^{\frac{1}{2}}\left(\alpha_1+\alpha_2+\alpha_3+\alpha_1^{3}\alpha_2^{3}\alpha_3^{3}\right)+\ldots+t^2\left(\#\right)+\ldots+t^{4}\,,
\end{equation}
where 
\begin{equation*}
\#=1+\alpha_1^{3}\alpha_2+2\alpha_1^2\alpha_2^2+\alpha_1\alpha_2^{3}+\alpha_1^{3}\alpha_3+\alpha_1^2\alpha_2\alpha_3+\alpha_1\alpha_2^2\alpha_3+\alpha_2^{3}\alpha_3+2\alpha_1^2\alpha_3^2
\end{equation*}
\begin{equation}
+\alpha_1\alpha_2\alpha_3^2+2\alpha_2^2\alpha_3^2+\alpha_1^{3}\alpha_2^{3}\alpha_3^2+\alpha_1\alpha_3^{3}+\alpha_2\alpha_3^{3}+\alpha_1^{3}\alpha_2^2\alpha_3^2+\alpha_1^2\alpha_2^{3}\alpha_3^{3}\,.
\end{equation}
In contrast to the previous case of $N=3$, we now have at order 2 many marginal operators. At the maximal order 4, we have the single operator $\left(X_1X_2X_3X_{4}\right)^2$ which is neutral under the discrete symmetries and can also be written in various different ways using the $F$-term relations. Therefore, we see that there are inequivalent exactly marginal operators such as $X_1X_2X_3X_{4}$ and $X_1^2X_2^2$ that have a nilpotency index 3, and exactly marginal operators such as $X_1X_2X_3^2$ and $X_1^{3}X_2$ that have a nilpotency index 2 (powers of which yield no contribution at order 4 due to  F-term relations). 
Since we have explored here just one special direction of the conformal manifold (which preserves $N-1$ copies of the $\mathbb{Z}_{N}$ symmetry), various different exactly marginal operators may have different nilpotency indices. This is entirely consistent with the general picture we have advocated for.

Finally, we discuss the general case of $N$ fields, with a conformal manifold $\mathcal{M}_c^{(N)}$. The number of marginal operators is
\begin{equation}
\dim \mathcal{M}_c^{(N)} =\frac{4^{N-1}\Gamma\left(\frac32 +N-1\right)}{\Gamma(N+1)\Gamma\left(\frac32\right)}-N^2\,,
\end{equation} 
which grows rapidly for large $N$. We now show that the nilpotency index of an exactly marginal operator in a generic direction and at a generic point on the conformal manifold is
\begin{equation}
k(\mathcal{M}_c^{(N)})=N-1\,.
\end{equation} 
Note that this nilpotency index saturates the bound \eqref{eq:bound_on_index}, and that it agrees with the explicit result found above for $N=3$ (when $N=4$ we have only calculated the nilpotency index along a specific direction in the conformal manifold, so we cannot infer $k(\mathcal{M}_c^{(4)})$ from the result).

In order to find $k(\mathcal{M}_c^{(N)})$, we should consider a generic point along the conformal manifold corresponding to
\begin{equation}
\label{dW}
\delta W=\sum_{I=1}^{\dim \mathcal{M}_c^{(N)}} h^I\,\mathcal{O}_I\,,
\end{equation}
where $\mathcal{O}_I$ are the exactly marginal operators. We should also consider a generic exactly marginal operator $\mathcal{O}=\sum_I c^I \mathcal{O}_I$ with $c^I$ some generic coefficients, and find its nilpotency index. 

As the Hilbert series corresponding to the general marginal deformation \eqref{dW} is given by \eqref{Hn} with $\alpha_j=1$, we still have a single operator at the maximal $R$-charge $2(N-2)$. As a result, we have $\mathcal{O}^{N-1}\sim 0$ which means that $k(\mathcal{M}_c^{(N)})\leq N-1$. We now argue that $\mathcal{O}^{N-2}\not\sim 0$, implying that $k(\mathcal{M}_c^{(N)})= N-1$. Expanding $\mathcal{O}^{N-2}=\left(\sum_I c^I \mathcal{O}_I\right)^{N-2}$, we find that it is a sum of terms with the maximal $R$-charge $2(N-2)$. However, as discussed above, there is a unique operator $M$ at this $R$-charge and therefore we get some function of the couplings $h_I$ multiplying this operator: 
\begin{equation}
\mathcal{O}^{N-2}=p(c^I,h^I)M\,, 
\end{equation}
where $p$ is a polynomial in the $c^I$'s whose coefficients are rational functions of the $h^I$'s. We therefore have $\mathcal{O}^{N-2}\sim 0$ only if $p(c^I,h^I)=0$, which happens only on some complex codimension 1 (or higher) subspaces.\footnote{Note that $p$ cannot be the zero polynomial, since for generic coefficients this occurs only if all possible products of $N-2$ of the marginal operators $\mathcal{O}_I$ vanish. However, we know that there exists a nonzero operator $M$ with maximal $R$-charge, which can be written as a sum of products of $N-2$ marginal operators. If all such products vanish then the maximal $R$-charge operator must vanish as well, which is a contradiction.}
Thus we find that the nilpotency index is $k(\mathcal{M}_c^{(N)})=N-1$ for this conformal manifold.

\

\subsection{$4d$ $\mathcal{N}=1$  Free Vector Multiplet}

Consider the $4d$ $\mathcal{N}=1$ supersymmetric $U(1)$ vector multiplet. 
This is a free theory which contains a photon along with a Weyl fermion (the photino). 
In this theory, the operator 
\begin{equation}
\mathcal{O}=W^\alpha W_\alpha\,,
\end{equation}
is exactly marginal, with the corresponding parameter being the gauge coupling. Though this exactly marginal coupling does not affect correlation functions in flat space, it does effect various partition functions on non-trivial spaces \cite{Witten:1995gf}.
Now, due to Fermi statistics we have $\mathcal{O}^2=0$ and so the nilpotency index in this simple case is 
\begin{equation}
k(\mathcal{M}_c)=2\,.
\end{equation}
Note that in this example, $\mathcal{O}^2=0$ holds identically and not just in the chiral ring.

\

\subsection{$4d$ $\mathcal{N}=2$ Theories}
\label{4dN2}

Let us examine a four dimensional $\mathcal{N}=2$ conformal manifold and consider it as an $\mathcal{N}=1$ manifold. In this theory there are eight supercharges $Q_{i\alpha}$ and $\overline{Q}_{i\dot{\alpha}}$, where $i$ is the $SU(2)$ $R$-symmetry index, that satisfy the usual supersymmetry algebra 
\begin{equation}
\label{N2alg}
\left\{ Q_{i\alpha},\overline{Q}_{j\dot{\alpha}}\right\} =2\delta_{ij}\sigma_{\alpha\dot{\alpha}}^{\mu}P_{\mu}\,\,\,\,,\,\,\,\,\left\{ Q_{i\alpha},Q_{j\beta}\right\} =\left\{ \overline{Q}_{i\dot{\alpha}},\overline{Q}_{j\dot{\beta}}\right\} =0\,.
\end{equation}
In order to consider this theory in $\mathcal{N}=1$ notation, we choose the supercharges of this $\mathcal{N}=1$ sub-algebra to be $Q_{1\alpha}$ while the supercharges $Q_{2\alpha}$ are the ones extending ${\cal N}=1$ to ${\cal N}=2$. Using this formulation, the generators $\mathcal{O}_{I}$ of the conformal manifold are as usual $\mathcal{N}=1$ chiral primary operators of dimension 3. However, in the $\mathcal{N}=2$ superconformal algebra they are actually descendants of Coulomb-branch ($\mathcal{N}=2$ chiral primary) operators $\Phi_I$ of dimension 2 \cite{Argyres:2015ffa,Cordova:2016xhm}, 
\begin{equation}
\label{OPhi}
\mathcal{O}_{I}=\left\{Q_2^{\alpha},\left[Q_{2\alpha},\Phi_{I}\right]\right\}\,.
\end{equation}
To see this, recall that in $\mathcal{N}=1$ theories the operators that are added to the Lagrangian are of the form $\left\{ Q_1^{\alpha},\left[Q_{1\alpha},\mathcal{O}_{I}\right]\right\} $ (along with their complex conjugates) while in $\mathcal{N}=2$ theories they take the form $\{Q_1^{\alpha},[Q_{1\alpha},\{Q_2^{\beta},[Q_{2\beta},\Phi_{I}]\}]\}$. 

Now, as every $\mathcal{N}=1$ preserving vacuum is also $\mathcal{N}=2$ invariant,\footnote{This property can be readily verified using the supersymmetry algebra \eqref{N2alg}. From $\{Q_{1\alpha},\overline{Q}_{1\dot{\alpha}}\}=2\sigma_{\alpha\dot{\alpha}}^{\mu}P_{\mu}$ we obtain the known result that a $Q_{1\alpha}$ preserving vacuum has zero energy. Then, since the same anti-commutation relation is satisfied also by $Q_{2\alpha}$, we get (using unitarity) that this zero-energy vacuum is also $Q_{2\alpha}$ invariant. There are interesting exceptions to this theorem in the case that supersymmetry is non-linearly realized~\cite{Hughes:1986fa,Ferrara:1995xi,Antoniadis:1995vb} but since here we are interested in SCFTs this does not affect our arguments.} the vacuum expectation values $\langle\mathcal{O}_{I}\rangle$ taken in $\mathcal{N}=1$ vacua vanish, as can be seen by substituting \eqref{OPhi} and using the invariance of the vacua under the action of $Q_{2\alpha}$. Therefore, the exactly marginal operators $\mathcal{O}_{I}$ are all nilpotent everywhere on the conformal manifold of an $\mathcal{N}=2$ theory. This means that there are no hypersurfaces on such manifolds where some of the operators $\mathcal{O}_{I}$ can cease to be nilpotent and have nonvanishing vacuum expectation values.
The computation of the nilpotency index however may be still nontrivial.

\subsection{$3d$ $\mathcal{N}=2$ Wess-Zumino Model with a Cubic Superpotential}

We consider the $3d$ $\mathcal{N}=2$ theory of three chiral multiplets $\Phi_i$ ($i=1,2,3$) discussed in \cite{Strassler:1998iz,Baggio:2017mas}, given by a canonical K\"ahler potential and a cubic superpotential as follows:  
\begin{equation}
W=\lambda_{1}\Phi_{1}\Phi_{2}\Phi_{3}+\frac{\lambda_{2}}{6}\left(\Phi_{1}^{3}+\Phi_{2}^{3}+\Phi_{3}^{3}\right).
\end{equation}
This model was discussed briefly in subsection \ref{elementary}, and its $2d$ version was discussed at length in subsection \ref{2d}.  In $3d$, this superpotential is relevant in the UV and the theory flows to an IR fixed point which is part of a one (complex) dimensional conformal manifold parameterized by $h=\lambda_{2}/\lambda_{1}$, taking values in $\mathbf{CP}^{1}$ \cite{Strassler:1998iz}.

Let us calculate the nilpotency index of the exactly marginal operator, denoted by $\mathcal{O}$. As shown in \cite{Baggio:2017mas} (and discussed in subsection \ref{2d}), at a general point on the conformal manifold the chiral ring consists of a finite number of operators. Explicitly, at $R$-charge $2/3$ we have the three fields $\Phi_i$, at $R$-charge $4/3$ there are three quadratic relevant operators, and at $R$-charge $2$ the sequence terminates with the single exactly marginal operator $\mathcal{O}$. There are no more operators at higher orders, and we therefore have $\mathcal{O}^{2}\sim0$ at a generic point, implying that 
\begin{equation}
k(\mathcal{M}_c)=2\,.
\end{equation}
As discussed above, this nilpotency index only applies to generic points, and can jump along complex codimension 1 lines. Indeed, the exactly marginal operator is no longer nilpotent at the point $\lambda_2=0$ but it becomes nilpotent as soon as we turn a nonzero $\lambda_2$.\footnote{
	Let us note that the WZ theory discussed here with $\lambda_2=0$ is known as the $XYZ$ model and it has an IR dual description in terms of ${\cal N}=2$ $U(1)$ gauge theory with a charge $+1$ field $Q$ and a charge $-1$ field $\widetilde Q$ \cite{Aharony:1997bx}. This gauge theory has two $U(1)$ global symmetries. The $U(1)_A$ symmetry under which $Q$ and $\widetilde Q$ have the same charge $+\frac12$, and the topological $U(1)_J$ symmetry with the conserved current being $\epsilon_{\mu\nu\rho}F^{\mu\nu}$. Under the topological symmetry the two monopole operators, $V_+$ and $V_-$, have opposite charges and both have charge $-\frac12$ under the $U(1)_A$. Our calculation of the nilpotency index can be viewed as a prediction for the dual theory.}

\

\subsection{$4d$ $\mathcal{N}=4$ $SU(2)$ SYM}
\label{N4SU2}

Consider the conformal manifold of the four-dimensional $\mathcal{N}=4$ SYM theory with gauge group $SU(2)$. Using $\mathcal{N}=1$ notations the theory contains a vector superfield in the adjoint representation of the gauge group with the corresponding field-strength chiral multiplet $W_{\alpha}$, and a scalar chiral multiplet $\Phi$ transforming in the adjoint representation of $SU(2)$ and in the fundamental representation of an $SU(3)$ global symmetry. Denoting the representations of these groups by $(\boldsymbol{R}_{SU(2)},\boldsymbol{R}_{SU(3)})$, we have $W_{\alpha}$ transforming in the $(\boldsymbol{3},\boldsymbol{1})$ and $\Phi$ in the $(\boldsymbol{3},\boldsymbol{3})$. 

The theory also has a superpotential given by 
\begin{equation}
\label{cub}
W=h\left(\Phi^{3}\right)_{\left(\boldsymbol{1},\boldsymbol{1}\right)}\,,
\end{equation}
and $\mathcal{N}=4$ supersymmetry implies that $h$ and the (complexified) gauge coupling are related such that a linear combination of $(\Phi^{3})_{\left(\boldsymbol{1},\boldsymbol{1}\right)}$ and the glueball operator 
\begin{equation}
\label{S}
S=-\frac{1}{32\pi^{2}}\left(W^{\alpha}W_{\alpha}\right)_{\left(\boldsymbol{1},\boldsymbol{1}\right)}
\end{equation}
is the exactly marginal operator parameterizing the one dimensional conformal manifold. However, there is a chiral ring relation due to the Konishi anomaly (which is the supersymmetric version of the ABJ anomaly) \cite{konishi1984anomalous,konishi1985functional} that can be used to simplify the form of this operator. Let us consider the anomalous conservation of the $U(1)$ symmetry rotating the $\Phi$ fields with charge $1/3$, 
\begin{equation}
\overline{D}^{2}J_{U(1)_{\Phi}}=h\left(\Phi^{3}\right)_{\left(\boldsymbol{1},\boldsymbol{1}\right)}-4S\,.
\end{equation}
Since the LHS is a $\overline{Q}$-exact term it is zero in the chiral ring and we get the relation
\begin{equation}
\label{phiSrel}
h\left(\Phi^{3}\right)_{\left(\boldsymbol{1},\boldsymbol{1}\right)}-4S\sim 0\,.
\end{equation}
Therefore, the exactly marginal operator $\mathcal{O}$ can be chosen to be given by the glueball superfield alone, 
\begin{equation}
\mathcal{O}=S\sim \frac{1}{4}h\left(\Phi^{3}\right)_{\left(\boldsymbol{1},\boldsymbol{1}\right)}\,.
\end{equation} 

As discussed in subsection \ref{4dN2}, in four dimensional theories with at least $\mathcal{N}=2$ supersymmetry the exactly marginal operators are nilpotent since they are $Q$-descendants of Coulomb-branch operators. Therefore, $\mathcal{O}$ is nilpotent and we would next like to find its nilpotency index for generic $h$. In order to do this, we first employ an observation made in \cite{Cachazo:2002ry} according to which $S^N$ (for an $SU(N)$ gauge group) is classically $\overline{Q}$-exact and so is zero in the chiral ring (classically), while $S^{N-1}$ is not. It is then natural to ask if and how, the classical relation $S^2\sim0$ in the theory we consider, is modified quantum mechanically. If this relation does not receive corrections, it would result in the nilpotency index of $\mathcal{O}$ being 2. 

To investigate the possible quantum corrections, we notice that they can only take the following form, 
\begin{equation}
\label{SP}
S^{2}+f\left(h\right)P\sim 0\,,
\end{equation}
where $f(h)$ is a holomorphic function that vanishes at the free point of the conformal manifold (corresponding to the classical limit) and $P$ is a scalar chiral operator of $R$-charge 4 that is a singlet of the $SU(3)$ global symmetry. It is important to comment that we do not consider operators $P$ which are given by $S^2$, $((\Phi^{3})_{\left(\boldsymbol{1},\boldsymbol{1}\right)})^{2}$ or $S\,(\Phi^{3})_{\left(\boldsymbol{1},\boldsymbol{1}\right)}$ (or a combination of them) since it would then follow from \eqref{phiSrel} and \eqref{SP} that $S^2\sim 0$, implying no quantum correction.  

To check whether a chiral ring relation of the form \eqref{SP} can take place, let us find the candidate operators that can serve as the operator $P$. We first notice that using the definition of $W_{\alpha}$ it can be shown that for every chiral operator $X$ we have (see {\it e.g.} \cite{Cachazo:2002ry,Ceresole:1999zs,Seiberg:2002jq})
\begin{equation}
\label{WXrel}
W_{\alpha}X=\left[\overline{Q}^{\dot{\alpha}},D_{\alpha\dot{\alpha}}X\right\},
\end{equation}
where $D_{\alpha\dot{\alpha}}$ is the  covariant derivative. Then, taking $X$ in this identity to be the adjoint fields $\Phi$ and $W_{\beta}$ we obtain that in the chiral ring, $\left[W_{\alpha},\Phi\right]\sim0$ and $\left\{ W_{\alpha},W_{\beta}\right\} \sim0$. As a consequence of this the generators of the chiral ring are of the form $\textrm{Tr}\Phi^{k}$, $\textrm{Tr}\left(\Phi^{k}W_{\alpha}\right)$ and $\textrm{Tr}\left(\Phi^{k}W^{\alpha}W_{\alpha}\right)$ where Tr denotes a trace over the gauge indices ({\it i.e.} the operator is in the singlet of the gauge group). The various scalar operators of $R$-charge 4 that can be constructed from these generators are listed in the table below.\footnote{Notice that only independent operators are listed in this table. In particular, the $(\boldsymbol{1},\boldsymbol{1})$ representation of $\textrm{Tr}\left(\Phi W^{\alpha}\right)\textrm{Tr}\left(\Phi^{2}W_{\alpha}\right)$ is just the operator $\textrm{Tr}\left(\Phi^{3}\right)\textrm{Tr}\left(W^{\alpha}W_{\alpha}\right)$ already written in the table (note also that there are no possible quantum corrections to this relation which will result in an inequivalent relation).} 
\begin{center}
	\begin{tabular}{|c|c|}
		\hline
		Operator & Representations $\left(\boldsymbol{R}_{SU(2)},\boldsymbol{R}_{SU(3)}\right)$\tabularnewline
		\hline 
		\hline 
		$\textrm{Tr}\left(\Phi^{6}\right)$ & $\left(\boldsymbol{1},\boldsymbol{1}\right),\,\,\left(\boldsymbol{1},\boldsymbol{27}\right),\,\,\left(\boldsymbol{1},\boldsymbol{28}\right)$\tabularnewline
		\hline 
		$\textrm{Tr}\left(W^{\alpha}W_{\alpha}\right)\textrm{Tr}\left(W^{\alpha}W_{\alpha}\right)$ & $\left(\boldsymbol{1},\boldsymbol{1}\right)$\tabularnewline
		\hline 
		$\textrm{Tr}\left(\Phi^{3}\right)\textrm{Tr}\left(W^{\alpha}W_{\alpha}\right)$ & $\left(\boldsymbol{1},\boldsymbol{1}\right)$\tabularnewline
		\hline 
		$\textrm{Tr}\left(\Phi^{2}\right)\textrm{Tr}\left(\Phi W^{\alpha}W_{\alpha}\right)$ & $\left(\boldsymbol{1},\boldsymbol{8}\right),\,\,\left(\boldsymbol{1},\boldsymbol{10}\right)$\tabularnewline
		\hline 
		$\textrm{Tr}\left(\Phi W^{\alpha}\right)\textrm{Tr}\left(\Phi^{2}W_{\alpha}\right)$ & $\left(\boldsymbol{1},\boldsymbol{8}\right)$\tabularnewline
		\hline
	\end{tabular}
	\label{tab}
\end{center}
The representation $\left(\boldsymbol{1},\boldsymbol{1}\right)$ of $\textrm{Tr}\left(\Phi^{6}\right)$ corresponds to the operator $((\Phi^{3})_{\left(\boldsymbol{1},\boldsymbol{1}\right)})^{2}$ and we therefore see that all the singlets under the global symmetry which appear in the table come from  $((\Phi^{3})_{\left(\boldsymbol{1},\boldsymbol{1}\right)})^{2}$, $S^2$ or $S\,(\Phi^{3})_{\left(\boldsymbol{1},\boldsymbol{1}\right)}$. As a result, we conclude that there is no operator $P$ in the theory that can appear in a relation of the form \eqref{SP} and so the operator $S^2$ remains a trivial element of the chiral ring also quantum mechanically for generic $h$.\footnote{At certain isolated points in the complex $h$ plane there can in principle be a decomposition of the multiplet in which $S^2$ decomposes into  short multiplets \cite{Cordova:2016emh,Kinney:2005ej,Dolan:2002zh,Gadde:2010en} such that $S^2$ becomes a nontrivial element of the chiral ring.} Therefore, $\mathcal{O}^2\sim0$ and the nilpotency index is
\begin{equation}
k(\mathcal{M}_c)=2\,.
\end{equation}

\

Let us make a brief comment about $SU(N)$ $\mathcal{N}=4$ SYM with $N>2$. Viewed as an $\mathcal{N}=1$ model this is a theory of three adjoint chiral superfields $\Phi_i$ along with the superpotential 
\begin{equation}
W=h\,f_{abc}\Phi_{1}^{a}\Phi_{2}^{b}\Phi_{3}^{c}\,,
\end{equation}
(and the usual gauge kinetic term such that $h\sim \tau$ as required by $\mathcal{N}=4$ supersymmetry) where $f_{abc}$ is the antisymmetric invariant of the gauge group. As in the case of $SU(2)$ gauge group (and as discussed in subsection \ref{4dN2}), the exactly marginal operator ${\cal O}=f_{abc}\Phi_{1}^{a}\Phi_{2}^{b}\Phi_{3}^{c}$ that appears in the superpotential is nilpotent (it is plausible that the nilpotency index is $N$ but we do not provide a proof here). However, unlike the case of $SU(2)$ discussed above, this theory has two additional $\mathcal{N}=1$ preserving exactly marginal operators. The full chiral ring depends in a complicated manner on the position on the conformal manifold \cite{Berenstein:2000ux}. But, following our  arguments, the exactly marginal operators must all be generically nilpotent. It would be nice to understand this in detail.

\

\subsection{$4d$ $SU(3)$ SQCD with $N_f=9$}

Our last example is the four dimensional $\mathcal{N}=1$ $SU(3)$ SQCD with nine flavors.
This theory was first considered in \cite{Leigh:1995ep} and  it was shown that it has a seven-dimensional conformal manifold in \cite{Green:2010da}. The $SU(3)$ SQCD with nine flavors  can be connected to various geometric constructions related to compactifications of minimal $SU(3)$ $6d$ SCFT \cite{Razamat:2018gro} as well as the rank one E-string theory \cite{Razamat:2019vfd} on Riemann surfaces.
In particular in the former context it can be obtained by compactifying the $6d$ SCFT on a sphere with $10$ punctures \cite{Razamat:2018gro} with the seven exactly marginal deformations corresponding to the complex structure moduli of this surface.

Here  we will focus on the complex one-dimensional locus of the conformal manifold  discussed in \cite{Leigh:1995ep}. On this locus, the theory has an $SU(3)^6$ continuous global symmetry. The matter content is given by fundamental fields $Q_i$ and antifundamental fields $\widetilde{Q}_{i}$ ($i=1,2,3$) each  transforming in the $\bf3$ of an $SU(3)$ global symmetry group. Denoting the representations under the various groups by, 
\begin{equation}
\left(\boldsymbol{R}_{SU(3)_{\textrm{Gauge}}};\boldsymbol{R}_{SU(3)_{i}};\boldsymbol{R}_{SU(3)_{\tilde{i}}}\right)\,\,\,\,\,\,\,\left(i=1,2,3\right)\,,
\end{equation}
the superpotential can be written as follows,
\begin{equation}
W=h\sum_{i=1}^{3}\left[\left(Q_{i}^{3}\right)_{\left(\boldsymbol{1};\boldsymbol{\vec{1}};\boldsymbol{\vec{1}}\right)}+\left(\widetilde{Q}_{i}^{3}\right)_{\left(\boldsymbol{1};\boldsymbol{\vec{1}};\boldsymbol{\vec{1}}\right)}\right]\,,
\end{equation}
where $\boldsymbol{\vec{1}}\equiv(\boldsymbol{1},\boldsymbol{1},\boldsymbol{1})$ and $h$ parameterizes the direction that we are considering.

A special locus on the conformal manifold is the zero-coupling point where the theory is free. At this point the symmetry is $U(9)\times U(9)$ and we would like to examine its breaking as we move away from this point. We first notice that at nonzero coupling both of the $U(1)$ factors are broken (where one combination of them is non-anomalous). Next, let us focus on the breaking of the $SU(9)$ acting on the $Q_{i}$'s (the same applies also to the other $SU(9)$ acting on the $\widetilde{Q}_{i}$'s). It is broken to the subgroup $SU(3)^{3}$ and the broken off-diagonal currents, given by 
\begin{equation}
J_{ij}=\bar{Q}_{i}e^{V}Q_{j}\,\,\,\,\,\,\left(i\neq j\right),
\end{equation}
recombine with the baryons $B_{ij}=\left(Q_{i}^{2}\right)Q_{j}$. Both $J_{ij}$ and $B_{ij}$ transform in the $\boldsymbol{\bar{3}}$ of $SU(3)_{i}$ and in the $\boldsymbol{3}$ of $SU(3)_{j}$. Explicitly, away from the free point we have, 
\begin{equation}
\bar{D}^{2}J_{ij}=hB_{ij}\,,
\end{equation}
and as a result $B_{ij}$ are zero in the chiral ring. In addition to these 54 broken currents, there are 2 more broken currents corresponding to the 2 generators that are in the Cartan of $SU(9)$ and do not belong to the Cartan of either of the three preserved $SU(3)$ subgroups. Their non-conservation equations are as follows,
\begin{equation}
\bar{D}^{2}J_{U(1)_{1}}=h\left[\left(Q_{1}^{3}\right)_{\left(\boldsymbol{1};\boldsymbol{\vec{1}};\boldsymbol{\vec{1}}\right)}-\left(Q_{2}^{3}\right)_{\left(\boldsymbol{1};\boldsymbol{\vec{1}};\boldsymbol{\vec{1}}\right)}\right],
\end{equation}
\begin{equation}
\bar{D}^{2}J_{U(1)_{2}}=h\left[\left(Q_{2}^{3}\right)_{\left(\boldsymbol{1};\boldsymbol{\vec{1}};\boldsymbol{\vec{1}}\right)}-\left(Q_{3}^{3}\right)_{\left(\boldsymbol{1};\boldsymbol{\vec{1}};\boldsymbol{\vec{1}}\right)}\right],
\end{equation}
resulting in the baryons $B_{i}=(Q_{i}^{3})_{(\boldsymbol{1};\boldsymbol{\vec{1}};\boldsymbol{\vec{1}})}$ being equal in the chiral ring. Employing the anomalous non-conservation (or Konishi anomaly) equation of the $U(1)$ factor of $U(9)$ we obtain that these baryons are equivalent in the chiral ring to the glueball operator $S$ (defined in \eqref{S}). 
The same arguments also apply to $\widetilde{B}_{ij}$ and $\widetilde{B}_{i}$, yielding in total
\begin{equation}
B_{ij}\sim\widetilde{B}_{ij}\sim0\,\,\,\,,\,\,\,\,B_{i}\sim\widetilde{B}_{i}\sim h^{-1}S\,.
\end{equation}
As a result, the exactly marginal operator corresponding to the $h$ direction is just the glueball operator $S$. Moreover, employing the identity \eqref{WXrel} for $X=Q_{i},\widetilde{Q}_{i},W_{\beta}$ we obtain that in the chiral ring 
\begin{equation}
W_{\alpha}Q_{i}\sim W_{\alpha}\widetilde{Q}_{i}\sim\left\{ W_{\alpha},W_{\beta}\right\}\sim 0\,,
\end{equation}
and it is therefore generated by the following operators, 
\begin{equation}
\label{ChiR}
S\,,\,\,\,M_{ij}=Q_{i}\widetilde{Q}_{j}\,,\,\,\,B=\left(Q_{1}Q_{2}Q_{3}\right)_{\left(\boldsymbol{1};\boldsymbol{3},\boldsymbol{3},\boldsymbol{3};\boldsymbol{\vec{1}}\right)}\,,\,\,\,\widetilde{B}=\left(\widetilde{Q}_{1}\widetilde{Q}_{2}\widetilde{Q}_{3}\right)_{\left(\boldsymbol{1};\boldsymbol{\vec{1}};\boldsymbol{3},\boldsymbol{3},\boldsymbol{3}\right)}\,,
\end{equation}
where $M_{ij}$ is in the $\boldsymbol{3}$ of $SU(3)_{i}$ and in the $\boldsymbol{3}$ of $SU(3)_{\tilde{j}}$.

We would next like to show that the exactly marginal operator $S$ is nilpotent, and find its nilpotency index. In order to do so, we proceed along the same line of arguments as in subsection \ref{N4SU2} and consider the operator $S^3$, which is classically zero in the chiral ring, in the full quantum theory. As discussed in subsection \ref{N4SU2}, the only possible quantum correction to the classical relation $S^3\sim 0$ is given by 
\begin{equation}
\label{S3PQ}
S^{3}+f\left(h\right)P\sim 0\,,
\end{equation}
where $f(h)$ is a holomorphic function that vanishes at the free point and $P$ is a scalar chiral operator of $R$-charge 6 that is not given by $S^3$ and is a singlet of the global symmetry. If, in turn, no such operator $P$ exists in the theory the correction \eqref{S3PQ} cannot take place and we obtain $S^3\sim 0$ also in the quantum theory (for generic $h$), resulting in $S$ being nilpotent with the corresponding nilpotency index bounded by $3$.

We will next show that there is no such an operator $P$ in the theory. To do this, we notice that since the chiral ring is generated by \eqref{ChiR} and second and third powers of $B$ and $\widetilde{B}$ (as well as products with $S$ and powers of $M_{ij}$) do not contain singlets of the global symmetry, candidate operators for $P$ can only be given by 
\begin{equation}
S\left(M_{ij}^{3}\right)_{\left(\boldsymbol{1};\boldsymbol{\vec{1}};\boldsymbol{\vec{1}}\right)}\,.
\end{equation}
However, due to the relations,\footnote{Note that there are no possible quantum corrections to \eqref{M3} which will result in inequivalent relations.}
\begin{equation}
\label{M3}
\left(M_{ij}^{3}\right)_{\left(\boldsymbol{1};\boldsymbol{\vec{1}};\boldsymbol{\vec{1}}\right)}=\left(Q_{i}^{3}\right)_{\left(\boldsymbol{1};\boldsymbol{\vec{1}};\boldsymbol{\vec{1}}\right)}\left(\widetilde{Q}_{j}^{3}\right)_{\left(\boldsymbol{1};\boldsymbol{\vec{1}};\boldsymbol{\vec{1}}\right)}=B_{i}\widetilde{B}_{j}\sim h^{-2}S^{2}\,,
\end{equation}
we find that all these candidates are equivalent to $S^3$ in the chiral ring, and that there is no operator $P$ in the theory. Therefore, the exactly marginal operator $S$ is nilpotent for generic $h$ with the nilpotency index bounded by $3$.

\

%%%%%%%%%%%%%%%%%%%%%%%%%%%%%%%%

\section*{Acknowledgments}
The authors would like to thank O.~Aharony, A.~Schwimmer, J.~Song,  L.~Tizzano and S.~Yankielowicz for useful discussions. Z.K is supported in part by the Simons Foundation grant 488657 (Simons Collaboration on the Non-Perturbative Bootstrap).
The research of S.S.R. and O.S. is supported in part by Israel Science Foundation under grant no. 2289/18, by I-CORE  Program of the Planning and Budgeting Committee. OS is also supported by the Clore Scholars Programme. Z.K., O.S., and S.S.R. are also supported in part by BSF grant no. 2018204.
A.S. is supported by an Israel Science Foundation center for excellence grant and by the I-CORE
program of the Planning and Budgeting Committee and the Israel Science Foundation (grant number 1937/12). S.S.R. and A.S. are grateful to the SCGP for hospitality when this project was conceived.
\vfill\eject

%%%%%%%%%%%%%%%%%%%%%%%%%%%%%%%%

%\appendix

%%%%%%%%%%%%%%%%%%%%%%%%%%%%%%%%
\bibliographystyle{./aug/ytphys}
\bibliography{./aug/refs}
%%%%%%%%%%%%%%%%%%%%%%%%%%%%%%%%

\end{document}